\documentclass{JAC2000}

\usepackage{graphicx}

\begin{document}

\title{TOWARDS RELIABLE ACCELERATION OF HIGH-ENERGY AND HIGH-INTENSITY
ELECTRON BEAMS}

\author{
K. Furukawa\thanks{ e-mail: \texttt{<kazuro.furukawa@kek.jp>} }
\ and \ Linac Commissioning Group\thanks{
Linac Commissioning Group: 
N. Akasaka, A. Enomoto, \mbox{J.~Flanagan,}
H. Fukuma, Y. Funakoshi, K. Furukawa, T. Ieiri, N. Iida, \mbox{T.~Kamitani,}
M.~Kikuchi, H. Koiso, T. Matsumoto, S. Michizono, T. Nakamura,
Y.~Ogawa, S. Ohsawa, K. Oide, 
Y. Onishi, K. Satoh, M. Suetake and T.~Suwada} \\
High Energy Accelerator Research Organization (KEK) \\
Oho 1-1, Tsukuba, Ibaraki, 305-0801, Japan }

\maketitle

\begin{abstract} 

The KEK electron linac was upgraded to 8 GeV for the KEK B-Factory (KEKB)
project.  During commissioning of the upgraded linac, even while
continuing SOR ring injections, we had achieved a primary electron
beam with 10-nC $(6.24 \times 10^{10})$ per bunch up to 3.7-GeV for
positron generation.  This could be classified as one of the brightest
S-band linac's.

Since the KEKB rings were completed in December 1998, those 3.5-GeV
positron and 8-GeV electron beams have been injected with excellent
performance.  Moreover, we have succeeded in switching among the
high-intensity beams for KEKB and beams for two SOR rings with
sufficient reproducibility.

After commissioning of the KEKB ring started, we launched a
project to stabilize the intensity and quality of the high-current
beams furthermore, and have accomplished it while investigating every
conceivable aspect.

% During this improvement, we have acquired valuable experiences on
% tolerance studies and stabilization technique of the timing and rf
% systems.  We also gained a knowledge on the physical phenomena of the
% beams particularly of an emittance growth.  They are indispensable for
% the design and construction of the next generation accelerators such
% as a linear collider, an FEL and an injector for super-high-luminosity
% machines.

\end{abstract}

\section{INTRODUCTION}

The KEK B-factory (KEKB) project started in 1994 to study CP-violation
in B-meson decays with an asymmetric electron-positron collider. The
performance of the experiment depends on the integrated luminosity of
KEKB, and hence the beam injection efficiency from the injector linac.

In order to achieve efficient full-energy injection, the original
2.5-GeV electron linac was upgraded up to 8 GeV, while enforcing
the acceleration gradient by a factor of 2.5 and by extending the length
of the facility by about 40\%.  Because of the site limit, two
linac's with 1.7-GeV and 6.3-GeV were combined using a 180-degree
bending magnet system to form a J-shape linac.  Also the primary
electron beam was designed to be 10 nC per bunch to produce 3.5-GeV
positrons with 0.64 nC.

The upgraded electron/positron linac has been commissioned since the
end of 1997, even while continuing injection to the Photon Factory (PF).  
We
had overcome many practical difficulties, and had already achieved
most of the designed beam 
parameters\cite{kekb-linac-epac2000,lin-com-pac99}.

% The physics experiment (Belle) at the KEK B-factory 
% (KEKB) started in 1999 

% The commissioning has overcome several challenging issues and 
% the linac is providing fairly stable beams with very high availability. 

However, to pursue the capability of the linac and KEKB to its utmost
limit, we still continue to improve the quality of the beams. 

\section{Commissioning}

The commissioning started at the end of 1997 using the first part of
the linac just before completion of the upgraded linac.  In order
to carry it, a task force called the linac commissioning group was
formed, in which 7 persons from the linac and 12 persons from the ring
participated.  This group later became a part of the whole KEKB
accelerator commissioning group.

The beam was operated at the linac local control room at the
beginning.  After completion of the KEKB rings the operation rooms
for the linac and the ring were merged with some computer network and
video switch preparations. Part of the operation log-book has been
recorded electronically to facilitate communication between local
engineers and remote operators.

\section{Stability and Reliability}

After commissioning of the KEKB ring started, we realized
that it was necessary to manipulate the beam delicately and continuously in
order to maintain the quality of the high-intensity beams for a long
term without degrading the injection performance.  Thus, we have
launched a project to stabilize the intensity and quality of the
high-current beams.

\subsection{High-Current Beam}

At the beginning of the commissioning it was necessary to make much
effort to transport a 10-nC electron beam on to the positron
generation target.  It was often difficult to maintain the beam for more than
one hour.  Otherwise, local bumps had to be made to cure beam
instabilities, which were caused by a fluctuation of the accelerator
equipment and transverse wake-fields.

Such difficulties, however, were gradually resolved after
understanding the sources of the instabilities through careful beam
studies, as surveillance systems were installed for rf systems and
other equipment\cite{rf-ical99}. Since the commissioning had started
before completing the whole linac, some part of the accelerator
equipment was not operated at the optimum condition. The largest
contributions to the instabilities came from many parameters in the
pre-injector section\cite{preinjector-epac2000}.
                                     
Thus, we had realized that it was important to study the tolerances of
beams to each parameter.  Table~\ref{tab-tor} shows some of those
results.

\begin{table}[htb]
\begin{center}
\caption{Tolerances of a 10-nC beam}
\vspace{2mm}

% \begin{tabular}{ l | r @{.} l }
% \hline
% \hline
% \textbf{Parameter}  & \multicolumn{2}{c}{\textbf{Tolerance}} \\
% \hline
% Gun high voltage    & $\pm$ 0&38 \% \\
% Gun timing          & $\pm$ 45& ps \\
% SHB1 (114MHz) phase  & $\pm$ 1&1 deg. \\
% SHB2 (571MHz) phase  & $\pm$ 1&3 deg. \\
% Buncher phase       & $\pm$ 1&7 deg. \\
% Buncher voltage     & $\pm$ 0&47 \%   \\
% Sub-booster-A phase & $\pm$ 3&5 deg. \\
% Sub-booster-B phase & $\pm$ 4&0 deg. \\
% \hline
% \end{tabular}

\begin{tabular}{ llllll }
\hline
\hline
\multicolumn{3}{c}{\textbf{Parameter}} &
\multicolumn{3}{c}{\textbf{Tolerance}} \\
\hline
% Gun high voltage    & $\pm$ 0.38 \% \\
\hspace*{2mm}& Gun high voltage &\hspace*{2mm} &
\hspace*{2mm}& $\pm$ 0.38 \%    &\hspace*{2mm} \\
& Gun timing          &&& $\pm$ 45. ps   &\\
& SHB1 (114MHz) phase &&& $\pm$ 1.1 deg. &\\
& SHB2 (571MHz) phase &&& $\pm$ 1.3 deg. &\\
& Buncher phase       &&& $\pm$ 1.7 deg. &\\
& Buncher power       &&& $\pm$ 0.47 \%  &\\
& Sub-booster-A phase &&& $\pm$ 3.5 deg. &\\
& Sub-booster-B phase &&& $\pm$ 4.0 deg. &\\
\hline
\end{tabular}

\label{tab-tor}
% \vspace{-2mm}
\end{center}
\end{table}

These tolerance values were obtained to keep 90\% of the maximum
beam current at the positron production target by changing only one
parameter around a good set of parameters. Software to find a 
correlation was used in order to acquire these
data\cite{comm-cont-ical99}. 

For a long, term each parameter may drift independently. If the room
temperature changes, most parameters may correlate with it. Thus,
while the above tolerance values are good references to consider the beam
stability, the parameters of the equipment have to be kept within much better
limits.

In order to stabilize the equipment parameters following the above guidelines,
stabilization software, which will be described later, was implemented
as well as hardware improvements.

After such a challenging effort, we achieved a primary electron beam
with 10-nC ($6.24 \times 10^{10}$) per bunch up to 3.7-GeV for
positron generation, without any loss at the 180-degree bending
system.  This could be classified as one of the brightest S-band
linac's.

\subsection{Four Beam Modes}

It was anticipated that it might degrade the performance of the linac
to switch beams between four injection modes. After a high-current
beam was achieved, we sometimes found that the beam parameters
were not optimal. Actually, the beam parameters in the four beam modes are
quite different, as shown in Table~\ref{tab-mod}.

\begin{table}[htb]
\small
\begin{center}
\caption{Beam Modes of the Linac}
\vspace{2mm}

\begin{tabular}{ lcccc }
\hline
\hline
& \multicolumn{2}{c}{\textbf{KEKB}} & \textbf{PF} & \textbf{PF-AR} \\ 
& \textbf{HER} & \textbf{LER} \\
\hline
Energy   & 8 GeV   & 3.5 GeV & 2.5 GeV & 2.5 GeV \\
Particle & $e^-$   & $e^+$   & $e^-$   & $e^-$ \\
Charge   & 1.28 nC & 0.64 nC & 0.2 nC  & 0.2 nC \\
         &         & (10 nC)\footnotemark \\
Repetition & 50 Hz & 50 Hz   & 25 Hz   & 25 Hz \\
Refill \\
\hspace*{2mm}
   Time  &1-2 min. &5-10 min.&3-5 min. &3-5 min. \\ 
\hspace*{2mm}
 Interval&1 - 2 hr.&1 - 2 hr.&24 hr.   &2 - 4 hr. \\
\hline
\end{tabular}

\label{tab-mod}
% \vspace{-2mm}
\end{center}
\end{table}
\footnotetext{3.7-GeV primary electron beam.}

The major challenging issues here were reproducibility of the beams in
one of four modes, the reliability of switching and the switching speed to
improve the integrated luminosity.

In this area, software to switch the beam modes had been developed since
the beginning of the commissioning. In order to accomplish the above tasks,
the software was refined, especially in the magnet initialization for
the reproducibility and in recovery of the equipment failures for the
reliability. It can even be re-configured easily in several
aspects by an operator. The details are described
elsewhere\cite{switch-feedback-linac2000}.

Using this enhanced software, the loss time caused by beam mode
switching was made negligible, and the beams became well reproduced
over frequent mode switches.  The switching time for the KEKB modes
became 90 to 120 seconds, which is acceptable. Thus, it is not a major
issue at the linac any more.

There are several plans for experiments that use high-energy electrons
in the linac. An example is the slow positron facility for solid-state
and particle physics\cite{slow-pos-linac98}. While the priorities of
these experiments are currently low, new beam modes for them may be
added to the routine operation if it is possible to solve new
switching issues.

% and the experiment of positron production enhancement by a channeling
% radiation. 

\subsection{Beam Feedback Loops}

Even with the efforts on beam stabilization and reliable beam mode
switching, it was sometimes necessary to tune the equipment parameters
delicately in order to maintain some beam parameters in the long
term. Only some experts could tune the beam and it took some time.

Simple feedback loops to limit energy fluctuations of the beams had been
installed since the beginning of the
commissioning\cite{lin-efb-ical99}. Also the same software was applied
to stabilize equipment parameters, as already described above. It
was also applied to stabilize the beam orbits.  More than 30 feedback
loops have been installed and are working depending on the beam
modes.  The details are described
elsewhere\cite{switch-feedback-linac2000}.

These feedback loops have improved short-term linac stability, and
have cured long-term drifts as well. 
% Normally we don't manipulate beams much.  

% \subsection{Pre-injector Section}

\subsection{Beam Optics}

In order to reproduce the beam well under different conditions, the
beam optics along the linac must be understood well. We have
investigated several aspects to find any discrepancy between the design and
the real optics. 

In order to measure the beam emittance
well, both the Q-magnet-scan method and wire scanners have been used depending
on the locations.
The errors in energy gain evaluations along the linac were not small,
unfortunately.  We are trying to refine it using a gain derived from
the rf measurement, beam energy measurement by an analyzer magnet and
a longitudinal wake-field estimation. 

Using such beam information, software systems were developed to match
the beam optics at the fixed energy\cite{optics-match-epac2000} and to
re-match the optics after a rf-power
re-configuration\cite{optics-correc-epac2000}.  They are used daily, 
although it does not
cover the whole linac yet, since we have several matching points along the
linac. 

The effect of the transverse wake-field is not small, especially with
high-intensity beams, and it degrades the beam emittance and
the stability. Evaluation and reduction of the wake-field effects have been
tried with some success\cite{wake-simu}.  
% However, it is not used in routine operation yet.  
Quadrupole wake-field effects were also
observed for the first time\cite{wake-quad-epac2000}.

\section{Operation Statistics}

With the help of the above-mentioned improvement, the linac operation has become
fairly reliable.  The total operation time in FY 1999 was 7296 hours,
which was greatly increased owing to full KEKB
operation\cite{kekb-linac-epac2000}.  The availability of the linac
for injection was 99.0\%, which have been much improved.

The average intensity of the positron in spring 2000 was 0.62 nC,
which is just less than the safety limit at the beam-transport line.

\section{More Challenges}

\subsection{Discharge in Accelerator Structures}

The discharge in the accelerator structures at sections A1 (buncher
and the first normal structure) and 21 (positron generator) became
severe in March, 2000, where the beam charge (and loss) is high and is
surrounded by solenoid coils. It was found that the discharge
frequently occurred near the trailing edge of the rf pulses.

Thus, the wave guides at these sections were re-arranged to shorten
the pulse width, and the rf-power was optimized to the improved beams
with lower voltage. Then, such a discharge decreased the rate to less
than once a day.

Since it is important to understand the phenomena deeply, a test stand
for such stations was built for investigating discharge
phenomena as well as for the conditioning of accelerator structures.

\subsection{Two Bunch Acceleration}

In order to double the positron beam charge, it is considered to have
two bunches in a linac rf pulse.  Because of the rf synchronization
scheme between the linac and the ring, those bunches have to be
separated by 96 ns at minimum\footnote{275th bucket in the linac and
49th bucket in the ring}.

A preliminary study was made on this two-bunch scheme, which produced
promising results on the energy compensation of the second bunch with
a careful rf-pulse timing control. The energy difference was estimated to
be 2.5\% for the 8-nC beam comparing the longitudinal wake-field with
the one for a low-intensity beam. Devices for this scheme are under
preparation.

\section{Conclusions}

In commissioning of the KEKB injector linac we have overcome
challenging issues and have accomplished a stabilization project,
investigating every conceivable aspect.  The linac is providing fairly
stable beams with very high availability.

During normal operation, operators rarely change the beam
parameters. Instead, software for beam-mode switching and feedback
loops takes care of them. Since the charge limit at the beam-transport
line induced by the safety reasons will be removed soon, the
performance of the linac may be more enhanced. 

Throughout this improvement, we obtained valuable experiences on tolerance
studies and stabilization technique of the timing and rf systems,
especially at the buncher section.  We also gained knowledge concerning
the physical phenomena of the beams particularly of an emittance
growth.  They are indispensable for the design and construction of
the next-generation accelerators, such as a linear collider, an FEL
and an injector for super-high-luminosity machines.

\end{document}